**Ultrafast x-ray diffraction probe of terahertz field-driven soft mode dynamics**

M. Kozina[1], T. van Driel[1], M. Chollet[1], T. Sato[1], J.M. Glownia[1], S. Wandel[1], M. Radovic[2,3], U. Staub[2], and M.C. Hoffmann[1]

[1]*Linac Coherent Light Source, SLAC National Accelerator Laboratory, Menlo Park, California 94025, USA*

[2]*Swiss Light Source, Paul Scherrer Institut, 5232 Villigen PSI, Switzerland*

[3]*SwissFEL, Paul Scherrer Institut, 5232 Villigen PSI, Switzerland*

**Abstract:** We use ultrafast x-ray pulses to characterize the lattice response of $SrTiO_3$ when driven by strong terahertz (THz) fields. We observe transient changes in the diffraction intensity with a delayed onset with respect to the driving field. Fourier analysis reveals two frequency components corresponding to the two lowest energy zone-center optical modes in $SrTiO_3$. The lower frequency mode exhibits clear softening as the temperature is decreased while the higher frequency mode shows slight temperature dependence.

The development of high peak-field sources of few-cycle terahertz (THz) pulses[1] has enabled experiments exploring THz-driven excitations in solids. While optical measurements (e.g. transient reflectivity or absorption[2,3], second harmonic generation[4,5], Faraday rotation[6]) are commonly employed to interrogate the THz-induced dynamics, the results only indirectly reveal any structural perturbations. On the other hand, ultrafast x-ray sources including synchrotron slicing sources and x-ray free-electron lasers provide novel probes that can be used to explore structural dynamics via x-ray scattering[7–12]. The combination of single-cycle THz excitation with ultrafast x-ray diffraction probe pulses allows direct tracking of atomic displacements within the unit cell when driven by an intense electromagnetic field. Because the x-ray pulses are short compared to the carrier-envelope-phase-stable THz pulse, it is possible to study the sample response on a sub-cycle time scale while the driving field is still present.

Recently ultrafast THz fields have been proposed to drive domain switching in ferroelectric systems[5,13,14]. However, direct evidence of the concomitant ionic motion coupled to the domain flipping is lacking. Strontium titanate ($SrTiO_3$, STO) is a prototypical perovskite that is prevented from undergoing a ferroelectric phase transition at low temperature because of quantum fluctuations[15,16]. The similar structure of STO to the bulk perovskite ferroelectrics $BaTiO_3$ and $PbTiO_3$[17] suggests that this system may be used as a model case to explore the structural changes induced under excitation with a THz field compared to those that exhibit equilibrium ferroelectricity. Moreover, STO has several zone-center infrared (IR)-active phonon modes[18] within the bandwidth of single-cycle table-top THz radiation sources[1] that can be driven resonantly by intense THz pulses. Thus STO provides an interesting case for probing field-driven structural dynamics.

We performed time-resolved x-ray diffraction measurements on a thin 100 nm film of STO pumped by single-cycle THz radiation. The x-ray diffraction measurements were performed at the XPP end station[19] of the Linac Coherent Light Source (LCLS) in monochromatic mode. The x rays were tuned to 9.5 keV (~1 eV bandwidth) and were 20 fs FWHM in duration at 120 Hz repetition rate with a 120 µm spot size. The arrival time of the x-ray pulses relative to the pumping THz radiation was corrected using a spectral encoding mechanism[20] so that the effective jitter between the x rays and THz was less than 50 fs. Our x-ray signal was recorded using an area detector (CSPAD 140K)[21]. All x-ray diffraction intensity measurements were collected at the top of the ($2\bar{2}5$) diffraction peak for the STO film and integrated over a 2D projection of reciprocal space on the detector at fixed sample position. We show a schematic of the scattering geometry in Fig. 1B.

We generated single-cycle p-polarized THz pulses at 120 Hz via optical rectification of 1.3 μm 50 fs pulses from an optical parametric amplifier (OPA) in DSTMS[22]. The OPA was pumped by 800 nm radiation from a Ti:Sapphire system (120 Hz, 25 mJ, 40 fs). The THz field was measured using electro-optic sampling (EOS) in a 100 μm GaP crystal at the sample position using as a probe a small fraction of the 800 nm light not used to pump the OPA. The peak THz field strength was 250±50 kV/cm and the central frequency was ~3 THz with significant bandwidth from 0.5-6.5 THz (see Fig. 2B). The THz beam was propagated in a dry-nitrogen environment except for a few cm of ambient air immediately before the sample to mitigate any THz absorption by water vapor. See Fig. 1A for a diagram of this setup.

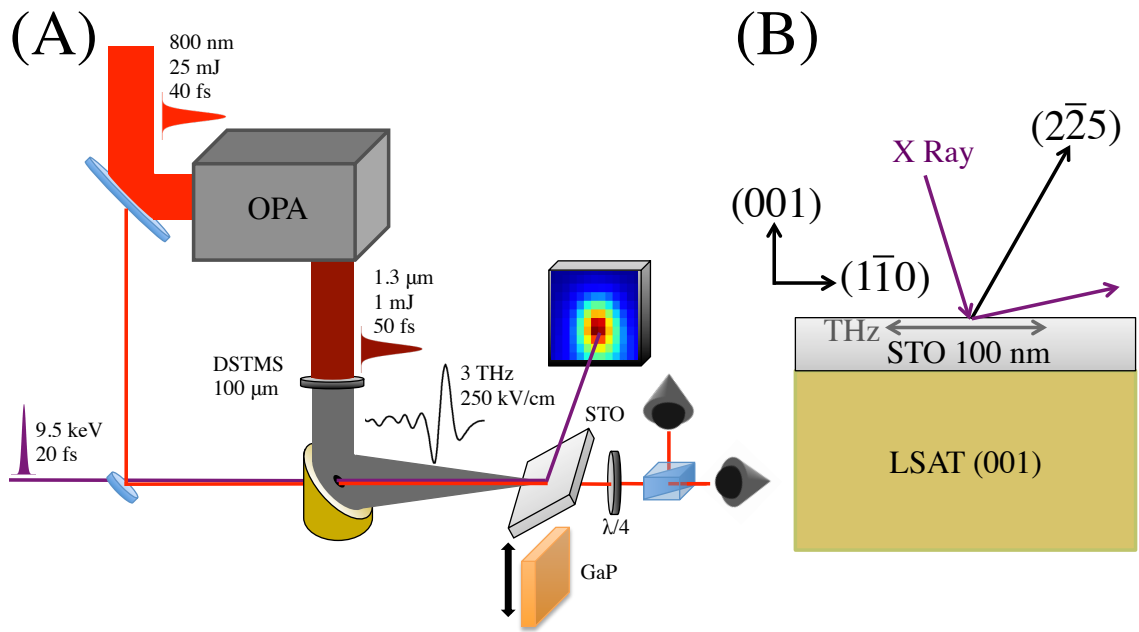

Fig. 1 (A) Schematic of experimental setup. Sample (SrTiO₃, STO) can be exchanged for GaP EOS crystal while maintaining interaction point. (B) Scattering geometry for $(2\bar{2}5)$ reflection. The THz polarization axis in the sample is shown in gray while the x-ray wavevector is in purple.

The times of arrival of the x-ray and 800 nm pulses were established by carrying out an 800 nm-pump, x-ray probe experiment on a bismuth thin film in the exact location of the STO sample and under otherwise identical conditions. On pumping with an 800 nm femtosecond laser pulse, Bi exhibits a structural change manifest as a rapid drop in the scattering intensity of the (111) diffraction peak[23]. We measured this signal and resolved the initial drop to a resolution of 50 fs. The relative time of arrival of the THz and 800 nm pulses was then chosen via EOS so that the peak of the THz field was coincident with the 800 nm pulse and hence the x-ray pulse to within our time resolution. This procedure allowed us to unambiguously compare the THz response of the sample with the incident THz field.

Our sample consisted of an epitaxial 100 nm STO film on a $(La_{0.3}Sr_{0.7})$ $(Al_{0.65}Ta_{0.35})O_3$ (LSAT) substrate with the (001) peak out of plane. For details of the sample growth see Ref. [24]. The sample temperature was tuned from 105 K to 320 K using a cooled nitrogen gas flow (Oxford Instruments Cryojet 5). The gas temperature provides a lower bound for the sample temperature, which is at most 10 K higher. Values quoted below correspond to the gas temperature.

In Figure 2A we show the fractional change in scattering intensity $\Delta I/I$ of the $(2\bar{2}5)$ Bragg peak of the STO film at 120 K as a function of time delay between the THz pump (black) and x-ray probe (blue). We define the fractional change in scattering intensity $\Delta I/I = [I(t)-I_0]/I_0$ where $I_0$ is the value of the scattered intensity before the THz pulse has arrived and $I(t)$ is the intensity at time delay $t$. Because the STO film is much thinner than the THz wavelength in the film, the permittivity of the LSAT substrate will dominate refraction effects. The relatively large value of the LSAT index of refraction in the THz regime[25,26] ensures that the axis of the THz polarization will lie completely within the film. In our scattering geometry, this is along the [1$\bar{1}$0] direction. The short wavevector of the THz radiation will couple only to zone-center optical modes, which will modulate the structure factor of the diffraction peak. We specifically chose the (2$\bar{2}$5) peak because its structure factor is particularly sensitive to ionic motion along the [1$\bar{1}$0] direction.

Overlaid with the x-ray diffraction data is the electric field of the THz pump (black) measured from EOS. We see a clear time-delay between the arrival of the THz field and the onset of structural changes in the STO manifest as a change in diffraction intensity. Moreover, while the initial decrease and then increase in scattering intensity follow roughly the THz waveform, there are persistent oscillations in the x-ray diffraction signal after the THz pulse has passed. We attribute these to excited zone-center optical phonons in the STO and describe in greater detail below.

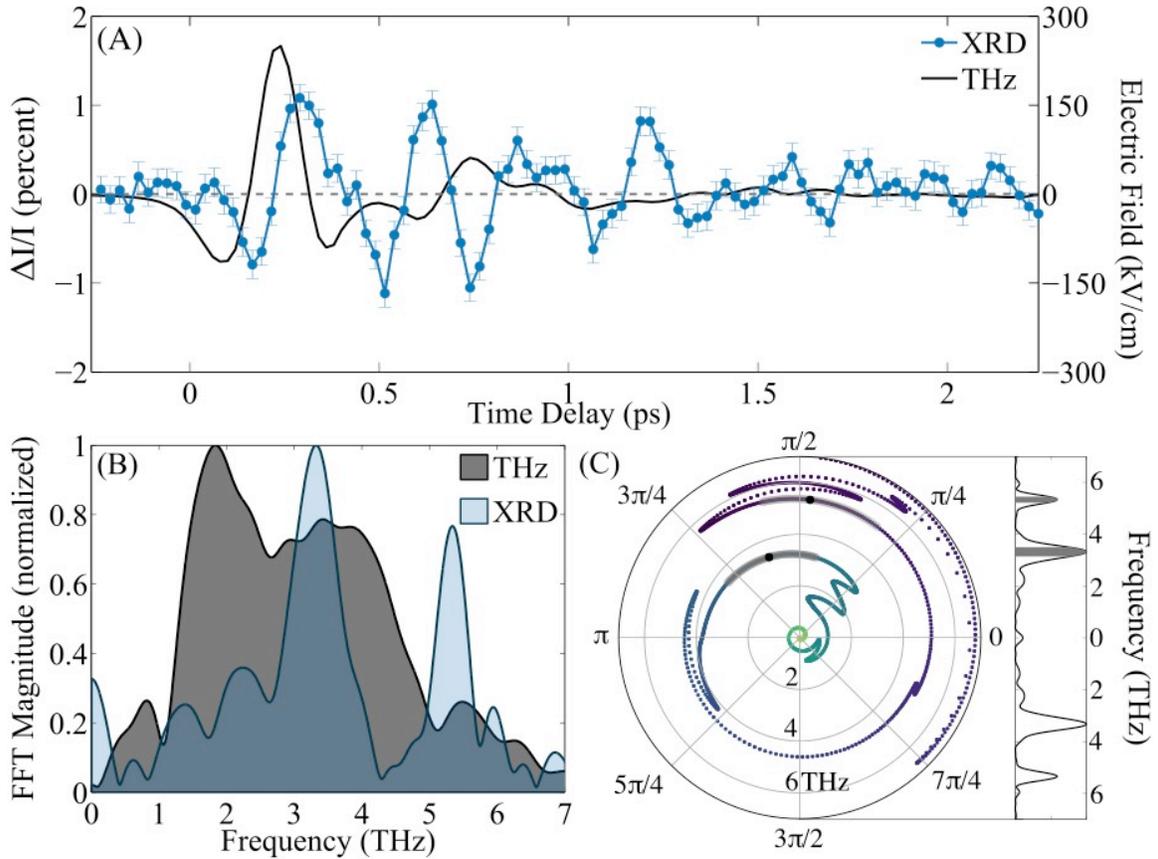

Fig. 2. (A) Electric field of THz excitation pump measured using electro-optic sampling (EOS) (black) and time-resolved change in ($2\bar{2}5$) peak intensity at 120 K (blue). Error bars on EOS data are smaller than line width. (B) Magnitude of the FFT of the data in (A) with zero-padding; THz (black) and x-ray diffraction (XRD) (blue). (C) Phase difference between the THz and XRD signals as a function of FFT frequency (radial coordinate). The gray shaded regions highlight the two peaks of the XRD FFT magnitude shaded in the right subpanel.

In Fig. 2B we show the magnitude of the Fast Fourier Transform (FFT) of the THz and time-resolved x-ray data. We utilized zero-padding in order to better resolve the phase change between the two signals, as shown in Fig. 2C. We observe two sharp peaks at frequencies consistent with known IR-active phonons at zone center[18], and label them $TO_1$ and $TO_2$. These peaks also explain the persistent oscillations in the x-ray scattering signal after the THz pulse has propagated out of the film. Because we are exciting on-resonance, we efficiently couple energy into both IR active modes, and so oscillations persist after the driving field has passed through the film. The phase difference between the THz pump and structural change is $\sim\pi/2$ (Fig. 2C) at each peak. This time delay between driving field and system response at resonance is to be expected for a driven damped harmonic oscillator model. The system response will delay the driving force by $\pi/2$ in agreement with our observations.

We assume that the THz couples only to the $TO_1$ and $TO_2$ modes, and can estimate from the change in scattering intensity the amount of motion of the ions within the STO unit cell. The ionic motion will change the structure factor for the STO unit cell, and the fractional change in the square of the structure factor is equal to the fractional change in scattering intensity. Note we ignore heating effects that would create an additional slow overall decay of the scattering intensity (e.g. strain waves, Debye-Waller factor modulation). In Fig. 3A, we plot the expected change in scattering intensity resulting from motion along either the $TO_1$ (solid blue) or $TO_2$ (dashed red) phonon eigenvector polarized parallel to the THz field (along the $[1\bar{1}0]$ direction). The gray shaded region corresponds to the largest intensity changes we observe in our scattering measurements. We show diagrams of the two phonon eigenvectors along the $[1\bar{1}0]$ direction in Fig. 3B ($TO_1$) and C ($TO_2$)[27]. The cubic symmetry of STO suggests that any ionic motion away from equilibrium will serve to only increase the scattering intensity of the $(2\bar{2}5)$ peak, thus effectively rectifying the signal. However the finite imaginary contribution to the atomic scattering factors shifts the minimum to a non-zero displacement, enabling measurements for low ionic motion to remain in a linear regime. Larger displacements will lead to a non-linear regime in the diffraction measurement (independent of any sample nonlinearity) that can lead to harmonics of the oscillation frequency as we move towards the nonlinear portion of the parabola in Fig. 3A.

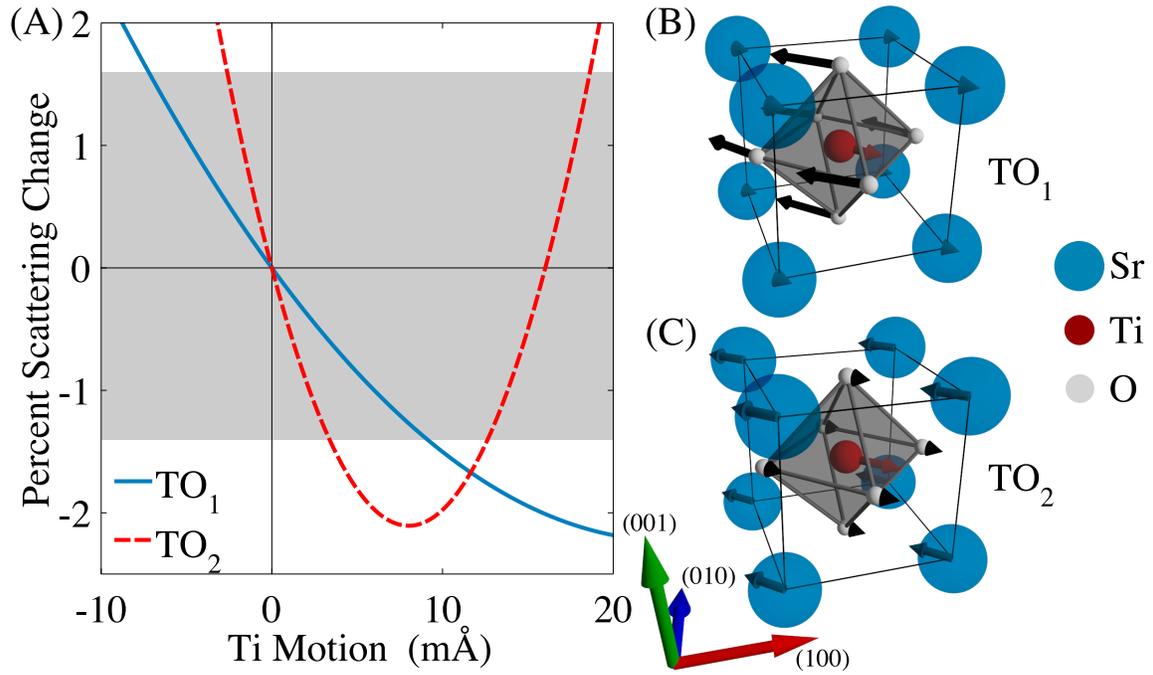

Fig. 3. (A) Calculated change in scattering intensity versus Ti motion along [1$\bar{1}$0] for two different eigenvectors, TO$_1$ and TO$_2$. The shaded patch in gray corresponds to the regime of changes in intensity we measure in Fig. 2A. (B) Cartoon of TO$_1$ eigenvector polarized along [1$\bar{1}$0]. Sr, Ti, and O are blue, red, and gray respectively. (C) Cartoon of TO$_2$ eigenvector polarized along [1$\bar{1}$0].

As a first approximation, if we assume the motion of the ions is along only the TO$_1$ phonon eigenvector, the maximum displacement of the Ti ion from equilibrium is about 0.01Å, or 0.25% of the lattice parameter, similar to values reported elsewhere via THz time-domain spectroscopy[2] (the estimated motion is ∼50% smaller assuming only TO$_2$ motion). This is about ten times smaller than the offset of the Ti ion in ferroelectric tetragonal BaTiO$_3$, which shifts along the ferroelectric polarization direction by 2% of the lattice constant[28]. The spectral weight of both phonon modes, however, is comparable (see Fig. 2B) and so the situation is more complex than a single-mode model admits.

Measurements of more Bragg peaks are required to further elucidate the structural dynamics of the STO cell because the diffraction measurement couples both $TO_1$ and $TO_2$ motions whether or not they interact in the sample.

In STO films on LSAT substrates the low-frequency soft mode undergoes reduced softening as a function of temperature[18,29] compared to bulk STO. To explore the change in coupling between the STO film and the THz pump, we tuned the sample temperature from 105 K to 320 K. In Figure 4A, we show the time-resolved change in scattering intensity as a function of temperature; in Figure 4B, we show the magnitude of the Fourier transform of this data (not padded), along with the square root of a fit of the power spectrum to two Gaussian peaks. Each spectrum exhibits two peaks, one that varies strongly with temperature and one that is nearly constant. The black lines are guides to the eye to highlight the temperature dependence of the central frequency of each peak. We identify the signal at the lower frequency as the soft mode $TO_1$, showing a clear reduction in frequency as the temperature is lowered. The other peak is close to the known value of the next zone-center IR active phonon mode $TO_2$ in STO and shows weak temperature dependence in agreement with IR measurements[18].

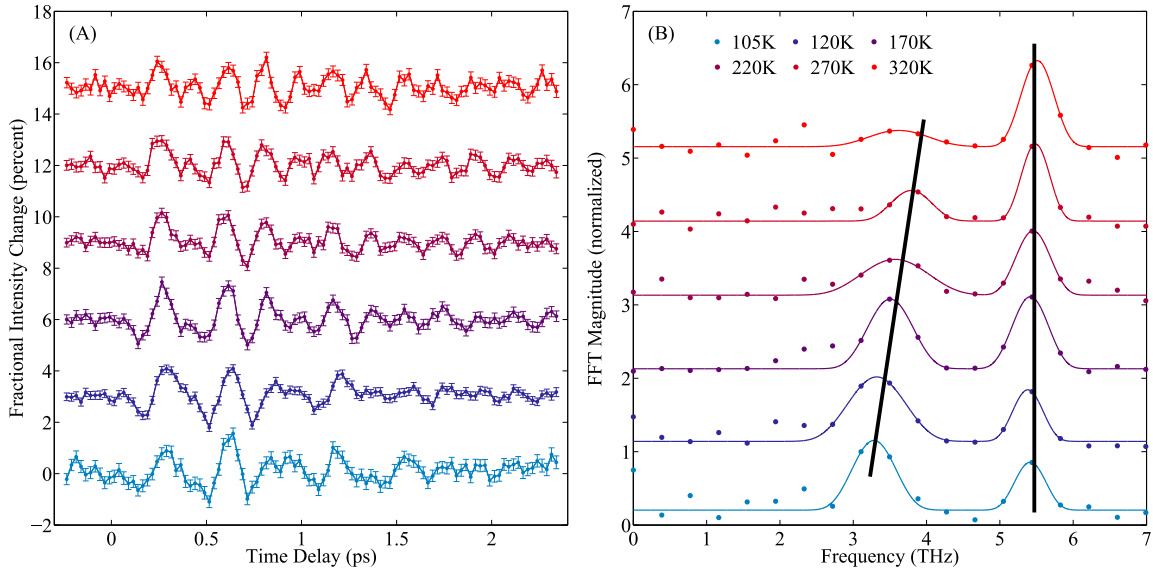

Fig. 4. (A) Time-resolved change in scattering intensity of the $(2\bar{2}5)$ peak after excitation with THz as a function of temperature. (B) Magnitude of Fourier transform of data in (A) overlaid with the square-root of a fit of the power spectrum to two Gaussian peaks. The dots are data and the solid lines are the resultant fit.

In Figure 5, we summarize the temperature dependence of the various fit parameters from the power spectra. Overlaid with our results (solid markers) are values from IR measurements taken on 107 nm STO films on LSAT from [18] (hollow markers). We find that the low frequency peak goes from 3.3 THz at 105 K to 3.80 THz at 270 K, in good agreement with values reported from IR reflectivity[18] and ellipsometry[29]. Moreover, the magnitude of the soft mode signal decreases with temperature while the $TO_2$ mode increases (Fig. 5B) even though the THz driving field spectral content is flat over the soft mode frequency range. A similar shift in spectral weight has been observed in hyper-Raman measurements of bulk STO at higher temperatures and was there attributed to coupling between the $TO_1$ and $TO_2$ modes[30].

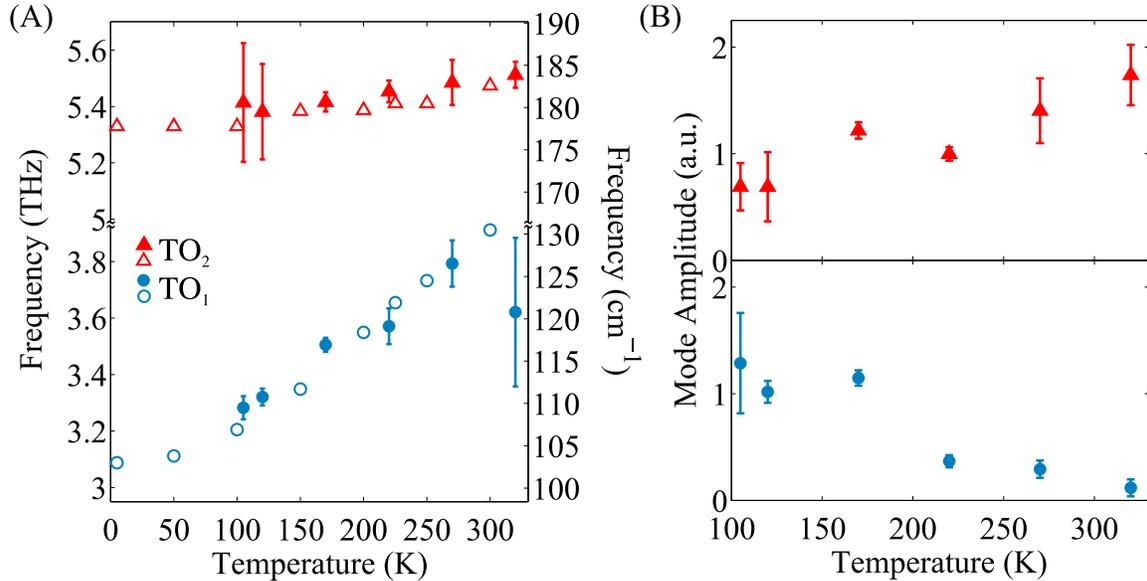

Fig. 5. Fit parameters from Fig. 4B as a function of temperature for each peak. The $TO_1$ mode (blue circles) is the low-frequency peak and the $TO_2$ mode (red triangles) is the high-frequency peak. Solid markers are from this work while hollow markers are from [18]. (A) Central frequency of peak. (B) Magnitude of the peak normalized to the peak value for $TO_1$ at 120 K. Error bars are 95% confidence intervals from the fitting routine.

The combination of excitation with single-cycle THz radiation and ultrafast x-ray diffraction expands the capabilities of THz spectroscopy. Because x-ray diffraction gives direct insight on the structural changes of a system, we can readily observe coupling between THz radiation and phonon modes. Using a THz-pump, x-ray probe measurement of STO we were able to directly observe the softening of the low-frequency mode as a function of temperature. Moreover, because our measurement was taken in the time domain, we were able to observe the phase shift between the THz field and the response of the STO system, reiterating the capability of time-domain measurements to study non-equilibrium processes as they happen.

Use of the Linac Coherent Light Source (LCLS), SLAC National Accelerator Laboratory, is supported by the U.S. Department of Energy, Office of Science, Office of Basic Energy Sciences under Contract No. DE-AC02-76SF00515. M.K. and M.C.H. are supported by the U.S. Department of Energy, Office of Science, Office of Basic Energy Sciences under Award No. 2015-SLAC-100238-Funding. U.S. acknowledges support from the National Center of Competence in Research, Molecular Ultrafast Science and Technology (NCCR MUST), a research instrument of the Swiss National Science Foundation (SNSF).